\documentclass[preprint,number]{elsarticle}
\usepackage[utf8]{inputenc}
\usepackage{amssymb}
\usepackage{graphicx}
\usepackage{epsfig}
\def\Figref#1{Fig.~\ref{#1}}   
\def\figref#1{Fig.~\ref{#1}}   
\def\eqref#1{(\ref{#1})}

\begin{document}
\title{Effect of boron dimers on the superconducting critical temperature in boron-doped diamond}
\author[mff,fzu]{B{\v r}etislav {\v S}op\'ik}
\ead{sopik@fzu.cz}
\author[mff,fzu]{Pavel Lipavsk{\' y}}
\address[mff]{Faculty of Mathematics and Physics, Charles University, Ke~Karlovu~3, 12116~Prague~2, Czech~Republic}
\address[fzu]{Institute of Physics, Academy of Sciences, Cukrovarnick{\' a}~10, 16253~Prague~6, Czech~Republic}

\begin{abstract}
We study how attractive boron correlations in boron-doped diamond affect 
the superconducting critical temperature. The critical temperature is obtained 
from the McMillan formula for strong coupling superconductors with the density 
of states evaluated in the dynamical cluster approximation. Numerical results 
for the cluster of $2\times 2\times 2$ atoms show that attractive correlations 
lower the density of states at the Fermi level. We argue that this might explain 
experimentally observed differences in critical temperatures of 100 and 111 
oriented films. 
\end{abstract}


\maketitle

\section{Introduction}

Diamond is a typical insulator with a band gap of $\sim 5.5 \,{\rm eV}$. It 
has been intensively studied for a~long time due to its compatibility with 
human tissues and unique physical properties like high transparency for a 
visible light or high thermal conductivity which are promising for future 
applications. When doped with boron, a \mbox{p-type} conductivity appears. As 
the doping concentration exceeds $\sim 4.5 \times 10^{21} {\rm cm}^{-3}$ the 
system becomes metallic and eventually superconducting, with an unexpectedly high 
critical temperature on the order of a few Kelvin. \cite{Ekimov04,Bustarret08,Blase09} 
The critical temperature depends on the method by which the sample was grown. 

Samples are most often prepared in thin layers, using the Microwave
Plasma-assisted Chemical Vapour Deposition (MPCVD) method with growth
orientations $100$ and $111$; see \cite{Klein07} and
\cite{Takano07} respectively. The $111$ samples have higher
$T_{\rm c}$ than $100$ samples.  There are also bulk samples prepared
with the High-Pressure High-Temperature (HPHT) method \cite{Yokoya05}
with $T_{\rm c}$ comparable to the $100$ samples. Some experimental
data are collected in \figref{fig4-11} below.

While the concentration of impurities is well under control in any of
the above preparation methods, a concentration of binary correlations between 
boron atoms is not and may cause the difference between samples. 
Here we discuss an effect of these correlations on $T_{\rm c}$.

Our study is motivated by previous experimental studies. From the Raman
scattering interpreted with the help of first-principle calculations 
Bourgeois {\em et al} suggested that boron atoms in boron-doped 
diamond form correlated multi-boron complexes, primarily dimers. \cite{Bourgeois06} 
This is confirmed by Mukuda {\em et al} who observed a strong NMR line 
of isolated boron impurities in $111$ samples, while this line is
rather weak in $100$ samples. \cite{Mukuda07} 

We will assume that boron atoms tend to form isolated dimers but not 
large clusters. This assumption is supported by {\em ab initio} investigations 
which show that the nearest-neighbour dimer is the most favourable of 
all two-boron configurations while an added third boron tends to remain
unassociated with the dimer \cite{Long08,Niu09}.

A dimer can be viewed as a molecule with bonding and anti-bonding
states made of symmetric and anti-symmetric combinations of bound
states of composing boron atoms. The symmetric state is deeper in the
band gap than the boron state, theqrefore it does not contribute to $N_{0}$, the
density of states at the Fermi energy. \cite{Goss06,Bourgeois06,Blase09} 
The energy of anti-symmetric states is above the isolated boron 
level and also does not contribute to $N_{0}$. The formation of 
dimers thus reduces $T_{\rm c}$.

To be able to perform numerically demanding configurational averaging
within the Dynamical Cluster Approximation (DCA), we use a model which excludes
many realistic features. We assume a simple cubic lattice with a single s-state 
per site, which does not cover the triple degeneracy of impurity state of 
boron \cite{Pogorelov05}. The impurity potential is restricted to a single 
site having no long-range Coulomb tails, and eventual screening of this 
potential is not assumed. This model was already studied by 
Shirakawa {\it et al} \cite{Shirakawa07} within the BCS theory, using the
Coherent Potential Approximation for disorder, and also studied by one of 
the present authors \cite{Sopik09} in the DCA for uncorrelated boron distribution. 

The DCA allows us to include statistical weight of boron complexes
given by their binary correlations. Nearest-neighbour positions of
two boron atoms are favoured by a dimerization energy. To avoid large clusters, 
we assume that once the atom is in a dimer it cannot gain the dimerization 
energy from another neighbouring atom. For this model we evaluate the density of 
states on the Fermi level as a function of boron concentration and dimerization
energy. The $T_{\rm c}$ is then obtained from the McMillan formula.

The paper is organized as follows. In section \ref{theory} we introduce the 
model Hamiltonian and describe the theory. In section \ref{dos} we calculate 
the density of states (DOS) and study the impact of boron correlations on 
$N_{0}$.  In section \ref{temperature} we compare our theoretical values 
of the critical temperature with experimental data. Section~\ref{conclusion} 
contains conclusions.

\section{Theory}
\label{theory}

The presence of boron impurities at concentrations reaching $5$\% of the
crystal atoms modify many properties of the crystal ranging from the
lattice constant, over the phonon spectrum and electron-phonon
coupling, to the dielectric function.  We neglect all these effects on
superconductivity and express material parameters of McMillan formula
as a function only of the density of states at the Fermi level
$N_{0}$. We introduce a model Hamiltonian and a method to calculate
the density of states. At the end of this section we specify the
probability distribution of boron clusters.

\subsection{McMillan formula}

The McMillan formula \cite{McMillan68} provides the critical temperature
\begin{equation}
	\label{McMillan}
	T_{\rm c} = \frac{\hbar\omega_{\rm D}}{1.45\,k_{\rm B}} \exp \Bigg \lbrack - 
	\frac{1.04\,(1 + \lambda)}{\lambda - \mu^{*}\big(1 + 0.62\,\lambda \big)} 
	\Bigg \rbrack.
\end{equation}
Since the disorder-dependence of the Debye frequency in the boron-doped diamond 
is weak, we use the pure-diamond value $\hbar\omega_{\rm D}/k_{\rm B} = 1860 \,{\rm K}$.

The screened pseudopotential
\begin{equation}\label{mu_star}
	\mu^{*}  = V N_{0} \left \lbrack 1 + V N_{0} \ln\frac{E_{\rm C}}{0.62 \,\hbar
	\omega_{\rm D}}\right \rbrack^{-1}
\end{equation}
is given by the strength of the Coulomb interaction $V$. We have found neither 
measurement nor estimate of $V$ in the literature, theqrefore we set this parameter 
from experimental $T_{\rm c}$. A single value $V = 3.6 \times 10^{-23}{\rm cm}^{3}{\rm eV}$
is used for all samples independently of boron concentration and correlation. 

We estimate the Coulomb cutoff energy as
\begin{equation} \label{cutoff}
	E_{\rm C} = \frac{1}{2} N_{0} \left ( \frac{\partial N_{E}}{\partial E} \biggl \vert_{E = 0} \right )^{-1}.
\end{equation}
For free particles this formula gives $E_{\rm C} \to \hbar^2 k_{\rm F}^2 / 2m$, 
usually used for the cutoff in metals \cite{VIK82}. In the impurity band the Fermi
momentum $k_{\rm F}$ is not well-defined, but expression \eqref{cutoff} in terms 
of the density of states is applicable. Since the Coulomb cutoff is only a subsidiary
quantity, we use its uncorrelated value also for 
correlated samples. The concentration dependence must be maintained as one
can see from the free-particle limit.

The electron-phonon coupling $\lambda$ we take from {\em ab initio}
calculations. The published results cover only a few concentrations, 
theqrefore we interpolate them in the spirit of the Morel-Anderson \cite{Morel61} formula
\begin{equation} \label{N_lambda}
	\lambda = \frac{U N_{0}}{1 + Q N_{0}}.
\end{equation}
Here $U$ represents a phonon-electron coupling strength and $Q N_{0}$ 
describes a screening. 

Using formulae \eqref{McMillan}-\eqref{N_lambda} we obtain $T_{\rm c}$
as a function of the density of states $N_{0}$ and its derivative
giving the Coulomb cutoff $E_{\rm C}$. The $N_0$ depends on the boron
concentration and dimerization energy. The $E_{\rm C}$ we take as a
function of only the concentration.

In \cite{Sopik09} it was shown that the Belitz theory with neglected
vertex corrections provides a better agreement with data on 100 samples 
than the McMillan formula. For impurity concentrations up to 5\% both 
approaches provide comparable results. In this paper we thus use the 
McMillan formula which is the standard approximation.

\subsection{Density of states}

Now we specify a model from which we obtain the density of states  
of non-interacting electrons as a function of the boron concentration 
$n_{\rm B}$. It is given by a Hamiltonian of the valence 
band in diamond $\hat {\mathcal H}_0$ and a random potential of boron 
impurities $\hat{ \mathcal V}$,
\begin{equation}\label{hamilt}
\hat{\mathcal H} = \hat{\mathcal H}_0 + \hat {\mathcal V} = \hat{\mathcal H}_0 + 
\sum_{i} \eta_{i}\delta \, \hat a^{\dagger}_{i} \hat a_{i},
\end{equation}
where $\eta_i=1$ at impurity sites and zero elsewhere, 
and $\delta$ is the potential amplitude. 

We employ the DCA for clusters of $2 \times 2 \times 2$ atoms on a cubic 
lattice. We only briefly introduce the DCA here; all details can be found 
in a previous paper of one of the authors \cite{Sopik09} or in the method's 
original paper \cite{Jarrell01}. The DCA provides a Green function 
averaged over all possible configurations of impurities 
on a cluster embedded in an effective medium. The key point is how to close the 
self-consistency constructing the effective medium from the averaged Green 
function.

Na\"ive constructions based on a tight-binding representation of the self-energy
have turned out to violate analytic properties and thus causality. In the DCA 
the self-energy is approximated in the momentum representation. The Brillouin 
zone is according to the size of the cluster divided into subzones -- in our 
case $2 \times 2 \times 2$ subzones. Within these subzones the self-energy is 
momentum-independent; that is, the selfenergy is represented by eight complex 
functions of frequency. 

All subzones contribute to the bare density of states
\begin{equation}
	\rho^{0}(E) = \sum_{\textbf{K}} \rho^{0}_{\textbf{K}}(E) ,
\label{denstat}
\end{equation}
where $\textbf{K}$ is a subzone index. In general, the subzone contribution 
$\rho^{0}_{\textbf{K}}$ is obtained by integrating over the subzone with the 
electron dispersion of the valence band. For simplicity we approximate 
$\rho^{0}_{\textbf{K}}$ in each subzone ${\bf K}$ by a semielliptical function
\begin{equation}\label{semielipt}
	\rho^{(0)}_{\textbf{K}}(E) = \frac{1}{N_{\rm c}}
	\frac{2}{\pi \, w_{\textbf{K}}} 
	\sqrt{1 - \bigg ( \frac{E - E_{\textbf{K}}}{w_{\textbf{K}}} \bigg )^{2} } ,
\end{equation}
where $N_{\rm c}$ is number of subzones and
\begin{eqnarray}
	w_{\textbf{K}} & = \frac{1}{2} \big(E^{{\rm max}}_{\textbf{K}} - 
	E^{{\rm min}}_{\textbf{K}}\big) , \\
	E_{\textbf{K}} & = \frac{1}{2} \big(E^{{\rm max}}_{\textbf{K}} + 
	E^{{\rm min}}_{\textbf{K}}\big) ,
\end{eqnarray}
reproduce the maximum $E^{{\rm max}}_{\textbf{K}}$ and minimum 
$E^{{\rm min}}_{\textbf{K}}$ energy in the subzone ${\bf K}$.

The density of states \eqref{denstat} by definition maintains the width 
of the valence band. 
This approximation also yields a correct 
curvature at the edge; it correctly reproduces an effective mass of 
holes near the band edge. This feature is vital for a realistic description 
of the relatively shallow impurity state.

The parameters $w_{\textbf{K}}$ and $E_{\textbf{K}}$ are fitted to the
tight-binding cosine band of width $22 \, {\rm eV}$. The impurity potential
$\delta = 8.91\,{\rm eV}$ reproduces the single-impurity bound state energy 
$0.37 \,{\rm eV}$ above the valence band. 

Each boron atom releases a singe hole, theqrefore the density of holes equals
$n_{\rm B}$. From this density we determine the Fermi energy $E_{\rm F}$ 
\begin{equation}
n_{\rm B} = 2\int_{E_{\rm F}}^{\infty} \rho(E) \,dE . 
\end{equation}
Here the factor of two accounts for spin. After we find $E_{\rm F}$ we shift
the energy reference point so that $E_{\rm F}=0$ as it is usual in the theory of
superconductivity. 

Mukuda {\it et al} \cite{Mukuda07} argued that the concentration of holes 
can differ from the boron concentration due to boron atoms in neutral B-H 
complexes or interstitial positions. 
On the other hand Klein {\it et al} \cite{Klein07} 
found that the effective number of carriers deduced from Hall-effect 
measurements was much larger than the number of boron atoms in samples. 
Since reliable hole concentrations are not accessible, we make the assumption
that all samples are doped ideally.

\subsection{Boron correlations}

The exact form of boron correlations in the diamond remains a matter 
of discussion. We introduce their attractive correlation via an 
interaction energy $\varepsilon_{\rm dim}$ between boron atoms 
located at the nearest neighbour sites.

Let $\mathcal{I}$ be a configuration of $n_\mathcal{I}$ boron atoms 
on $N$ sites of the cluster with $q_\mathcal{I}$ boron pairs in nearest neighbour
sites. Its probability 
\begin{equation}\label{distribution}
	p_\mathcal{I} = \frac{1}{Z}\, x^{n_\mathcal{I}}\,
(1-x)^{N-n_\mathcal{I}}\,{\rm e}^{-q_\mathcal{I}\frac{\varepsilon_{\rm dim}}{k_{\rm B}T_{\rm p}}}
\end{equation}
is given by a configuration statistical factor and by total energy
scaled by the temperature of sample preparation $T_{\rm p}$. 
Typical $T_{\rm p}$ for $100$ and $111$ samples is about $1100 \,{\rm K}$.
The probability is normalized to unity, $\sum_{\mathcal{I}}p_\mathcal{I}=1$, 
which determines $Z$. The parameter $x$ determines to the boron concentration
\begin{equation}
  n_{\rm B} = \sum_{\mathcal{I}}\,n_\mathcal{I}\,p_\mathcal{I}.
\end{equation}
For uncorrelated boron atoms, $\varepsilon_{\rm dim}\to 0$, it equals the boron 
concentration, $x\to n_{\rm B}$. In our case $N=8$ which allows for $2^8$
configurations. By symmetry one can reduce a number of configurations 
which must be treated numerically.

Apparently, averaging over small clusters underestimates contribution
of dimers. In the simplest case of $1\times 1\times 1$ cluster the
dimers are absent. In the assumed case of $2\times 2\times 2$ all
atoms are on the surface of the cluster with half of nearest neighbors
in the cluster and half outside. The outer neighbors are not accounted 
for which reduces the effect of dimers crudely by half. It can be 
partly compensated by an artificial increase of the dimerization 
energy by $\varepsilon_2=-k_{\rm B}T_{\rm p}\ln 2$. 

\section{Density of states in the presence of boron dimers}
\label{dos}

\begin{figure}
	\centerline{\includegraphics[height=8cm,angle=-90]{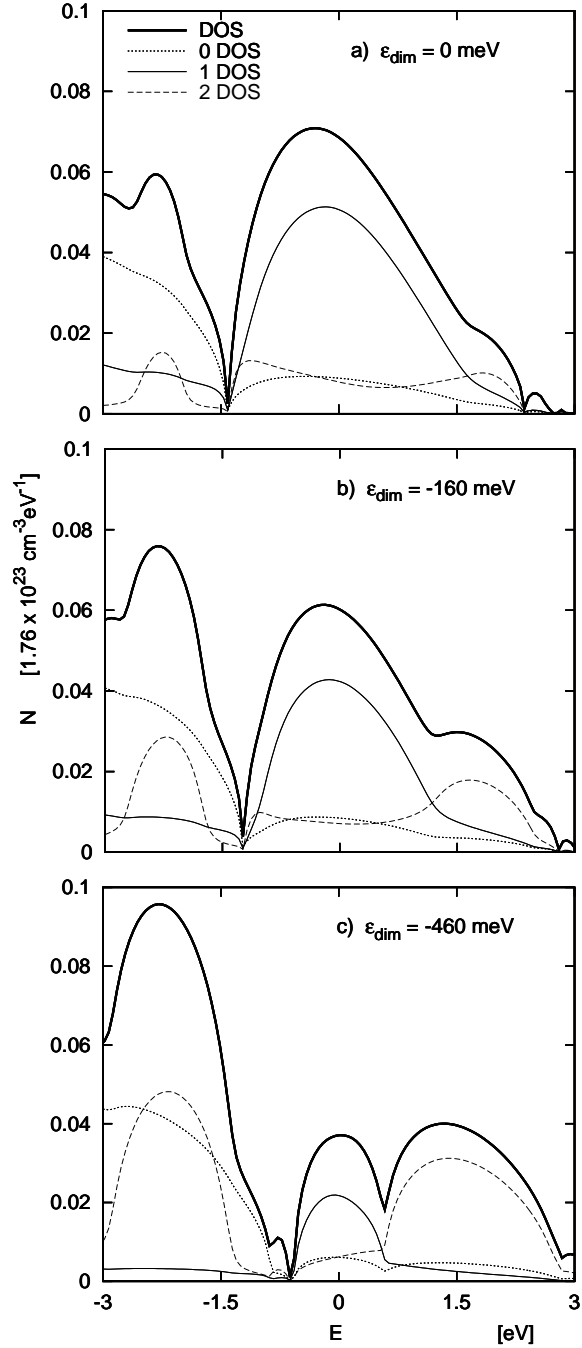}}
	\caption{\label{fig4-9} Density of states at doping $n_{\rm B} = 0.05$. 
	Positions of boron atoms at neighbour sites are a) uncorrelated 
	b) correlated by energy
	$\varepsilon_{\rm dim} = -160 \,{\rm meV}$ found from realistic 
	{\em ab initio} studies and c) correlated by $\varepsilon_{\rm dim} = -
	460 \,{\rm meV}$ fitted to give best agreement with experimental 
	data. The growth temperature is $T_{\rm p}=1100$~K.
	Thick line represents the DOS and thin lines its 
	decomposition into contributions according to number of boron 
	atoms in the $2\times 2\times 2$ cluster. 
	}
\end{figure}

Energy levels of a boron dimer are rather distinct from a level of 
an isolated atom. While a hole binds to an isolated atom at $0.37$~eV,
for our model the dimer has the symmetric level at $1.6$~eV and the 
anti-symmetric state forms only a resonant level in the valence band 
at $-2.2$~eV. For majority of samples the boron concentration is less
than $5$\%, theqrefore isolated boron atoms prevail in uncorrelated 
configurations. When the energy gain of two boron atoms at
neighbour positions exceeds the preparation temperature, 
$\varepsilon_{\rm dim}\gg K_{\rm B}T_{\rm p}\sim 94.8\,{\rm meV}$,
the fraction of dimers becomes large and this modifies the spectrum of
energies in the sample.

The effect of boron dimers on the overall density of states 
in the impurity band for $5$\% boron doping is shown in 
\figref{fig4-9}, where we compare the DOS for uncorrelated 
case a) with two correlated cases. In the case b) we take the boron-boron 
binding energy \mbox{$\varepsilon_{\rm dim}=-160\,{\rm meV}$} resulting from {\em ab initio}
calculations \cite{Bourgeois06,Long08,Niu09}. We refer to this
binding energy as a \emph{realistic correlation}. In the case c) we use
binding energy $\varepsilon_{\rm dim}=-460\,{\rm meV}$ which yields the best
fit of experimental data, see comparison in \figref{fig4-11}.
We shal refer to this binding ebergy as a \emph{fitted correlation}. 

It should be noted that {\em ab initio} studies compare energies of 
periodic crystals made of clusters B$_2$C$_{62}$ or B$_2$C$_{52}$ 
with selected configuration of boron atoms. Energies which one can
interpret as the binding energy of a nearest neighbour dimer achieve
values ranging from $-158\,{\rm meV}$, over $-205\,{\rm meV}$ to $-288\,{\rm meV}$
depending on authors and eventual gradient corrections. The 
fitted correlation energy $-460\,{\rm meV}$ is thus not far from these 
values.

We focus on the impurity band because the Fermi energy, $E_{\rm F} = 0$, lies there. 
The shape of the valence band is unimportant as it extends over 
energy interval from $-1.5\,{\rm eV}$ 
to $-24.5\,{\rm eV}$. The thick line represents the density of states and thin 
lines are its decomposition into contributions according to number 
of boron atoms in the $2\times 2\times 2$ cluster. All 
configurations of $n$ boron atoms are added together so that they contribute
with probability $p_n=\sum_{\mathcal{I}}^{n_\mathcal{I}=n}\,p_\mathcal{I}$.

Probabilities of $n$-boron clusters for the uncorrelated case are
$p_0=0.66$, $p_1=0.28$ and $p_2=0.05$. The zero-boron clusters contribute 
mainly to the valence band. In the impurity band they give a small 
contribution due to states tunneling from the surrounding medium. 
States from the single-boron clusters dominate the impurity band.

The two-boron clusters include 12 nearest-neighbour configurations 
which are the dimers, 12 from second-nearest neighbours, and four from third-nearest 
neighbours. For uncorrelated system in \figref{fig4-9}a)
the dimers increase the DOS at the upper edge of the impurity band around 
$E\sim 1.55$~eV, and in the valence band at $E\sim -2.15$~eV. 
In the correlated cases in \figref{fig4-9}b) and c) their contribution
is even more pronounced. The position of the two-boron levels are
shifted due to an increase of the dimer fraction in two-boron clusters
and due to feedback effect of surrounding medium. 

The small split band near $E=2.4$~eV in uncorrelated case in
\figref{fig4-9}a) is due to three-boron complexes. For realistic
correlations it shifts towards higher energies. For fitted
correlations it merges with the main impurity band. States of clusters
with more boron atoms are also present, but their contributions are
very small for the given concentration.

The realistic correlated case b) differs from the uncorrelated one a) mainly in
the probability of zero-, single-, and double-boron cluster contributions.
For given binding energy $\varepsilon_{\rm dim} = -160 \,{\rm meV}$ 
they are $p_0=0.70$, $p_1=0.21$ and $p_2=0.08$. In accordance with decreased $p_1$
and increased $p_2$, the contributions of double and higher boron states 
are enhanced and the peak associated with single boron states is reduced.
The lower probability of the single-boron clusters leads to the
decrease of the DOS at the Fermi energy. 

With fitted correlation c) the contributions from larger clusters of boron 
atoms become more apparent. For $\varepsilon_{\rm dim} = -460 \,{\rm meV}$ 
we find $p_0=0.77$, $p_1=0.07$ and $p_2=0.15$, theqrefore the number of
boron atoms in dimers is four times higher than the number
of isolated atoms. As one can see in \figref{fig4-9}c) the major part 
of the impurity band is due to dimer states. 

Note that even when dimers dominate, the Fermi energy remains in the
single-atom part of the impurity band. This follows from the fact that
each dimer binds two holes of spin up and down in the symmetric state.
Since the number of boron atoms in dimers equals the number of bounded
holes, two boron donors become passivated when they form a dimer. The
metallic band is thus formed only by unpaired boron atoms.

\begin{figure}
	\centerline{\includegraphics[height=8cm,angle=-90]{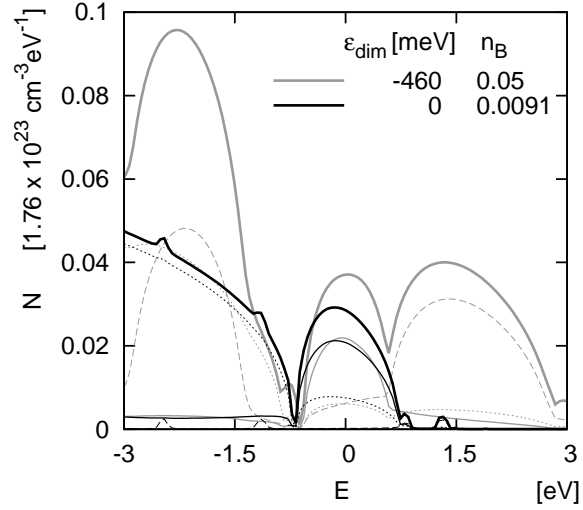}}
	\caption{\label{fig2} Approximation by subtracted dimers. The density 
	of states at doping $n_{\rm B} = 0.05$ and fitted correlation 
	$\varepsilon_{\rm dim} = -460 \,{\rm meV}$ (gray re-plot from \figref{fig4-9}c) 
	is compared with uncorrelated system of concentration $n_{\rm B}=0.0091$ (in
	black lines). Both systems have the same probability of single-boron
	clusters giving similar 1DOS (thin line). They have also similar 
	0DOS (dotted lines). The single-boron states, however, easily 
	tunnel into two-boron clusters giving appreciably larger 
	contribution of 2DOS (dashed lines) at the metallic band.  
	}
\end{figure}

The dimers influence the density of states in two ways. First, they 
effectively reduce the doping level.
Second, they contribute to the medium in which the band is formed. To 
demonstrate the second effect in 
\figref{fig2} we have included a DOS for uncorrelated system 
with $n_{\rm B}=0.0091$ giving the probability of single-atom clusters 
$p_1=0.07$ being equal to $p_1$ of the fitted correlation. One can see
that the approximation of a correlated system by a non-correlated one 
with dimers subtracted from the boron concentration leads to correct 
shape of the metallic band except for the contribution of clusters 
with dimers to the density of states at the Fermi level. For parameters 
in \figref{fig2} this contribution is a significant $\sim 20$\%,
assuming exponential dependence of $T_{\rm c}$ on $N_0$. 

\begin{figure}
	\centerline{\includegraphics[height=9cm,angle=-90]{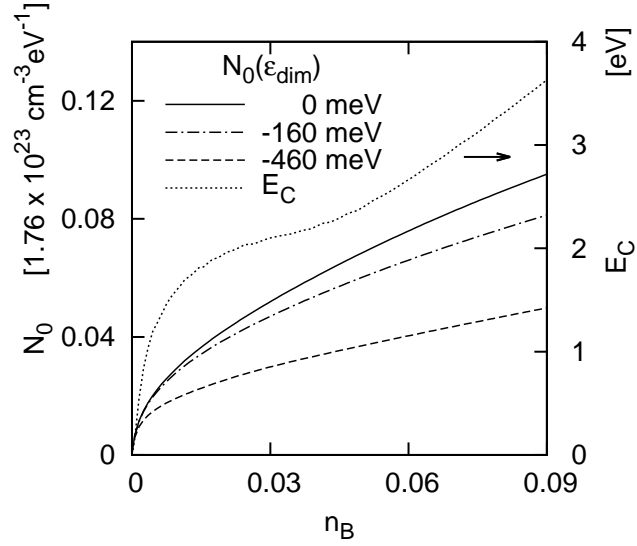}}
	\caption{\label{fig4-10}The density of states at the Fermi energy $N_{0}$ 
	as a function of boron concentration $n_{\rm B}$ for no (full line),
	realistic (dashed-dot line) and fitted (dashed line) correlation.
	The left vertical axis corresponds to the Coulomb cutoff $E_{\rm C}$
	(dotted line) evaluated for uncorrelated boron atoms.
	}
\end{figure}

The dependence of $N_{0}$ on the strength of boron correlations is shown
in \figref{fig4-10}. One can see that positive correlations in boron 
distribution lower the density of states at the Fermi level $N_{0}$ for any 
doping $n_{\rm B}$. \Figref{fig4-10} also presents the Coulomb cutoff
given by formula \eqref{cutoff}. This value is very close to the distance
of the Fermi level from the right band edge. An alternative definition
of the Coulomb cutoff from the bad edge, however, is imposed 
due to split off bands of trimers and larger clusters of low probability.

\section{Dependence of critical temperature on boron correlations}
\label{temperature}

Now we are ready to discuss the impact of correlations on 
$T_{\rm c}$. We ask the question whether different $T_{\rm c}$ of the 100 
and 111 samples can be explained by a different content of dimers. Mukuda 
{\em et al} \cite{Mukuda07} have observed a strong NMR line of single-boron
state in the 111 samples while this line was rather weak in the 100 samples.  
Based on this experimental fact we treat 111 samples as uncorrelated 
while in the 100 and HPHT samples we include dimers. 

\begin{figure}
	\centerline{\includegraphics[height=8cm,angle=-90]{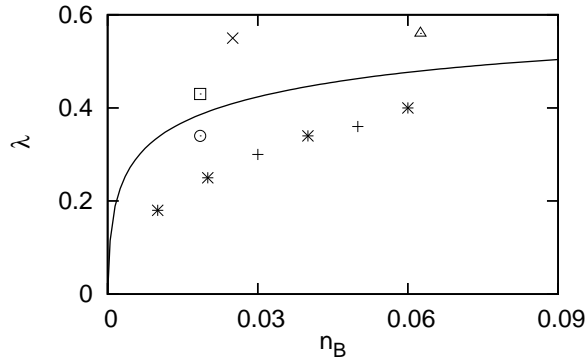}}
	\caption{\label{fig4-12} Coupling constant $\lambda$ as a function 
	of the boron concentration. Symbols represent results of {\em ab initio}
	calculations of~$\lambda$ for uncorrelated boron-doped diamond. 
	Solid line reproduces $\lambda$ dependence on $n_{\rm B}$ using 
	formula \eqref{N_lambda} with uncorrelated $N_{0}(n_{\rm B})$ 
	(see \figref{fig4-10}) and parameters 
	$Q = 35.7 \times 10^{-23}{\rm cm}^{3}{\rm eV}$, 
	$U = 23.3 \times 10^{-23}{\rm cm}^{3}{\rm eV}$. {\em Ab initio} results 
	were obtained within the Virtual Crystal ($+$ \cite{Boeri04}, 
	$\times$ \cite{Lee04}, $\ast$ \cite{Ma05}) and 
	Super Cell ($\square$ \cite{Blase04}, $\bigtriangleup$ 
	\cite{Xiang04}, $\bigcirc$~\cite{Giustino07}) 
	method.}
\end{figure}

First we specify material parameters. The $Q$ and $U$ are chosen to 
reproduce {\em ab initio} calculations 
of coupling parameter $\lambda$ in uncorrelated boron-doped diamond. 
Formula \eqref{N_lambda} with uncorrelated $N_{0}(n_{\rm B})$ and 
parameters $Q = 35.7 \times 10^{-23}{\rm cm}^{3}{\rm eV}$ and 
$U = 23.3 \times 10^{-23}{\rm cm}^{3}{\rm eV}$ is compared with {\em ab initio}
values in \figref{fig4-12}. Values $V$ and $\varepsilon_{\rm dim}$ are 
not accessible in print, we thus fit them to experimental values of 
$T_{\rm c}$. The best agreement is achieved for 
$V = 3.6 \times 10^{-23}{\rm cm}^{3}{\rm eV}$ and 
$\varepsilon_{\rm dim} = -460\, {\rm meV}$. 

\begin{figure}
	\centerline{\includegraphics[height=8cm,angle=-90]{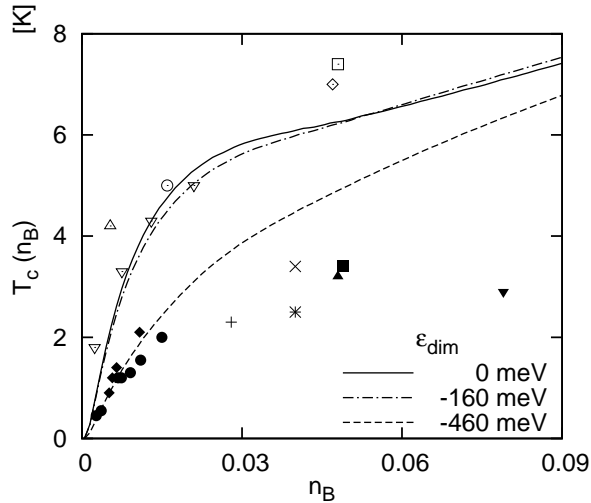}}
	\caption{\label{fig4-11} Critical temperature $T_{\rm c}$ as a function
	of boron doping for no (solid line), realistic (dashed-dot line), and 
	fitted correlation (dashed line). 
Open symbols represent 111 MPCVD data ($\square$~\cite{Takano07},
$\Diamond$~\cite{Yokoya05}, $\bigcirc$~\cite{Mukuda07},
$\bigtriangleup$~\cite{Takano04},
$\bigtriangledown$~\cite{Takano05}), full symbols 100 MPCVD data
($\bullet$~\cite{Klein07}, $\blacktriangle$~\cite{Takano07},
$\blacksquare$~\cite{Mukuda07}, $\blacklozenge$
\cite{Bustarret04}, $\blacktriangledown$ \cite{Umezawa05})
and crosses HPHT samples ($+$ \cite{Ekimov04}, $\times$
\cite{Sidorov05a}, $\ast$ \cite{Sidorov05b}).}
\end{figure}

In \figref{fig4-11} we compare $T_{\rm c}$ from McMillan formula
\eqref{McMillan} for uncorrelated
and correlated systems with experimental data for 111, 100 and HPHT samples. 
As one can see the realistic correlation energy $\varepsilon_{\rm dim}= -160\, {\rm meV}$ 
found in {\em ab initio} studies is too low to explain
differences in the critical temperature. With the fitted correlation
energy $\varepsilon_{\rm dim} = -460\, {\rm meV}$, the theory
reproduces trends of experimental values for small concentrations.
For $n_{\rm B}>0.02$ one finds only qualitative agreement.

Compared to the experimental data, the present study underestimates
the difference between 111 and 100 samples. It can be in part due to
our simple model, in part due to approximations employed to evaluate
the density of states. First of all, inside the $2\times 2\times 2$
cluster each atom has only three nearest neighbours while in the
crystal it has six. Accodingly, each boron can form only three dimers
inside the cluster while three dimers cross the cluster border. The
correlations corresponding to the former are covered in our treatment
while those corresponding to the latter are not.

There are also additional mechanisms that appear already in our model
but have not been included for simplicity of discussion. For example,
two boron atoms at second-neighbour positions form a symmetric state of
energy $0.91\,{\rm eV}$ which is close to the single-atom bound state and at
concentrations above $1$\% contributes to the mettalic band. The
anti-symmetric state of such a boron pair has energy at $-0.8\,{\rm eV}$ 
which forms a resonant level in the valence band and does not contribute to
the impurity band. One can crudely say one half of boron atoms in the
second-neighbour dimers are passivated.

\section{Conclusion}
\label{conclusion}

In this paper we have addressed the impact of correlations of boron
impurities on the density of states and thus on the critical
temperature $T_{\rm c}$ of the superconducting transition. To this end
we have calculated the impurity-band density of states for several
strengths of attractive correlations between boron atoms. Calculations
have been done within a one-band cubic lattice with on-site impurity
model employing the DCA method on a cluster of $2 \times 2 \times 2$
atoms.

We have found that correlations enhance states associated with
multi-boron complexes reducing the density of states at the Fermi
energy, see \figref{fig4-9} and \figref{fig4-10}. Following NMR data
we have approximated 111 oriented samples as an uncorrelated system
and 100 samples as a system with correlations. We have found that an 
{\em ab initio} binding energy $\varepsilon_{\rm dim} = -160 \,{\rm meV}$ of
boron dimer is not sufficient to explain differences in $T_{\rm c}$. 
From the best fit we have found the binding energy
$\varepsilon_{\rm dim} = -460\, {\rm meV}$ which explains experimental
data at least for boron concentrations below $2$\%.

The studied model has several shortcomings. It has overly-simplified
electronic band structure and the binding potential is single-site.
Also our treatment of the disorder covers only contributions of boron
complexes to the density of states but neglects other disorder effects
like the weak localization, which has been discussed as a possible
mechanism enhancing the superconductivity.~\cite{Mares07a}
Nevertheless, our results strongly indicate that binary
correlations are responsible for different critical temperatures of
the 111 and 100 samples.

Authors would like to thank Ji{\v r}{\' i} Mare{\v s} and
Jan Ma{\v s}ek for many useful discussions and also Peter Matlock 
for careful reading of a manuscript. The access to the METACentrum 
supercomputing facilities provided under the research plan MSM6383917201 
is also highly appreciated. This work was supported by research plans
MSM0021620834 and No.~AVOZ10100521, by grants, 
GA{\v C}R P204/10/0687, GA{\v C}R P204/10/0212 and DAAD project.

%

\end{document}